\documentclass[twocolumn,aps,pra,floatfix,10pt,groupedaddress,longbibliography]{revtex4-1}

\usepackage{xcolor}
\usepackage[utf8]{inputenc} 
\usepackage{bm}
\usepackage{amsmath,amsthm,amssymb}
\usepackage{graphicx} 
\usepackage{tikz,tkz-tab}
\usepackage[
  colorlinks=true,
  urlcolor=blue,
  linkcolor=blue,
  citecolor=blue
]{hyperref}

\begin{document}

\title{Relaxation and hysteresis near Shapiro resonances in a driven spinor condensate}

\author{Bertrand Evrard, An Qu, Karina Jim{\'e}nez-Garc{\'i}a, Jean Dalibard and Fabrice Gerbier}

\affiliation{Laboratoire Kastler Brossel, Coll{\`e}ge de France, CNRS, ENS-PSL Research University, Sorbonne Universit{\'e}, 11 Place Marcelin Berthelot, 75005 Paris, France}

\date{\today}

\begin{abstract}
We study the coherent and dissipative aspects of a driven spin-1 Bose-Einstein condensate (BEC) when the Zeeman energy is modulated around a static bias value. Resonances appear when the bias energy matches an integer number of modulation quanta. They constitute the atomic counterpart of Shapiro resonances observed in microwave-driven superconducting Josephson junctions. The population dynamics near each resonance corresponds to slow and non-linear secular oscillations on top of a rapid ``micromotion''. At long times and in a narrow window of modulation frequencies around each resonance, we observe a relaxation to asymptotic states that are unstable without drive. These stationary states correspond to phase-locked solutions of the Josephson equations generalized to include dissipation, and are analogous to the stationary states of driven superconducting junctions. We find that dissipation is essential to understand this long-time behavior, and we propose a phenomenological model to explain quantitatively the experimental results. Finally, we demonstrate hysteresis in the asymptotic state of the driven spinor BEC when sweeping the modulation frequency across a Shapiro resonance.
\end{abstract}

\maketitle	

\section{Introduction}
\label{sec:intro}

The Josephson effect is the hallmark of macroscopic quantum phenomena in quantum fluids, from superconductors\,\cite{josephson1962a,shapiro1963a,baroneJosephsonBook,kautz1996a} to superfluid Helium\,\cite{Pereverzev997a,avenel1999a,simmonds2001a,Sukhatme2001a}, polariton systems\,\cite{Sarchi2008,Shelykh2008,abbarchi2013} and ultracold atoms in double-well potentials\,\cite{albiez2005a,levy2007a,leblanc2010a,ryu2013a,valtolina2015a,pigneur2018a}. In all variants, the phase of a macroscopic wavefunction is controlled by an external bias parameter. In Superconducting Josephson Junctions (SCJJs), a voltage bias determine the relative phase between the two superconducting order parameter on each side of the junction and the supercurrent is proportional to the sine of this phase  \,\cite{josephson1962a,shapiro1963a,baroneJosephsonBook}. This leads to some remarkable behavior, as in the AC Josephson effect where a static voltage generates an oscillating current at the characteristic Josephson frequency $\omega_0$. Conversely, in the ``inverse AC Josephson effect''\,\cite{shapiro1963a,baroneJosephsonBook,kautz1996a} , an oscillating voltage near resonant with $\omega_0$ can carry a DC current across the junction. 

Multiple resonances occur when the drive frequency $\omega$ fulfills $k \omega=\omega_0$ for integer $k$\,\cite{shapiro1963a}. In SCJJs, these resonances appear in the form of Shapiro spikes in the voltage-current characteristics of a SCJJ junction driven at constant bias voltage, or steps at constant bias current\,\cite{kautz1996a}. Shapiro steps are at the core of Josephson voltage standards, which are essentially perfect frequency-voltage converters enabled by macroscopic quantum effects\,\cite{kautz1996a}.
Here, energy dissipation plays a crucial role\,\cite{kautz1996a}. Without dissipation, the system would not relax towards the exact resonance characterized by a phase locking of the macroscopic phase to the drive. 


Ultracold atoms offer a new situation, the so-called \textit{internal Josephson effect}, where coherent dynamics can occur between internal degrees of freedom \cite{leggett2001a,stamperkurn2013a}. Here we focus on the specific case of spin $F=1$ atoms, with $m_F$ the magnetic quantum number labeling the Zeeman components. An applied magnetic field plays the role of the external bias. The Josephson-like internal dynamics is generated by coherent, spin-changing collisions of the form $2\times  (m_F=0) \leftrightarrow (m_F=+1)+(m_F=-1)$ instead of single-particle tunneling \cite{law1998a,zhang2005a}. Most experimental studies of coherent spin-mixing dynamics were performed with only a static bias and no modulation\,\cite{chang2005a,zhang2005a,kronjaeger2005a,kronjaeger2006a,
black2007a,liu2009b,klempt2009a,klempt2010a}. Recent experiments explored the driven case, demonstrating the suppression of evolution by frequent ``kicks'' in spin space \cite{hoang2013a} and spin-nematic squeezing\,\cite{hamley2012a} near a parametric resonance\,\cite{hoang2016a}. Cold atoms variants of the Josephson effect (external\,\cite{albiez2005a,levy2007a,leblanc2010a,ryu2013a,valtolina2015a,pigneur2018a} or internal\,\cite{chang2005a,zhang2005a,
kronjaeger2005a,kronjaeger2006a,black2007a,liu2009b,
klempt2009a,klempt2010a}) occur with typical time scales on the order of milliseconds or longer, enabling a time-resolved study of the dynamics which is difficult to access in superconducting systems, where the microscopic time scales are in the picosecond range.


In this article, we report on the observation of the analogues of Shapiro resonances in a driven spin-1 BEC of sodium atoms. We characterize the resonances in a strongly non-linear regime, where the phase dynamics is not solely controlled by the external static bias.  We study not only the coherent dynamics at short times but also the relaxation at long times (tens of seconds, corresponding to thousands of oscillation periods). We find that the driven BEC relaxes to asymptotic states that are not stable without drive. In this sense, our system constitutes a many-body version of the celebrated Kapitza pendulum\,\cite{kapitza1951a,landaumec,citro2015a}. The stationary states correspond to phase-locked solutions of the Josephson equation generalized to include dissipation, analogous to the stationary states of driven SCJJs\,\cite{kautz1996a}.  

The paper is organized as follows. In Section\,\ref{sec:spinmixingnodrivin}, we review the main features of our experiment and of the theoretical description of spinor condensates. We highlight the analogies and differences with Josephson physics in superconducting junctions. We also discuss for later reference spin-mixing oscillations without driving, highlighting both the coherent features in Section\,\ref{subsec:spinmixingundriven}\,\cite{chang2005a,kronjaeger2005a,kronjaeger2006a,black2007a,liu2009b,klempt2009a,klempt2010a} and the dissipative aspects\,\cite{liu2009b} in Section\,\ref{sec:dissipationundriven}. In Section\,\ref{sec.ShortTimeDyn}, we turn to the driven system and characterize experimentally and theoretically the non-linear secular dynamics in the vicinity of the resonance. Measuring both the Zeeman population and the relative phase of the atoms, we identify two regimes, an ``oscillating regime'' where the atomic phase is phase-locked to the drive and a ``rotating regime'' where the atomic phase runs independently from the drive and wraps around a circle. We study the relaxation of the driven spin-1 BEC at long times in Section \ref*{sec:dissipation}. In a narrow frequency window around each Shapiro resonance, we observe relaxation to a non-equilibrium steady-state that has no analog in the undriven system. This allows us to discriminate between two phenomenological descriptions of dissipation used in the literature on Josephson-like models. While the two dissipative models are barely distinguishable from each other without driving, they differ spectacularly in the strongly driven case. We finally demonstrate that the system displays hysteresis near a Shapiro resonance in Section\,\ref{sec:hysteresis}, and conclude in Section\,\ref{sec:conclusion}.

\begin{figure}
	\centering
	\includegraphics[]{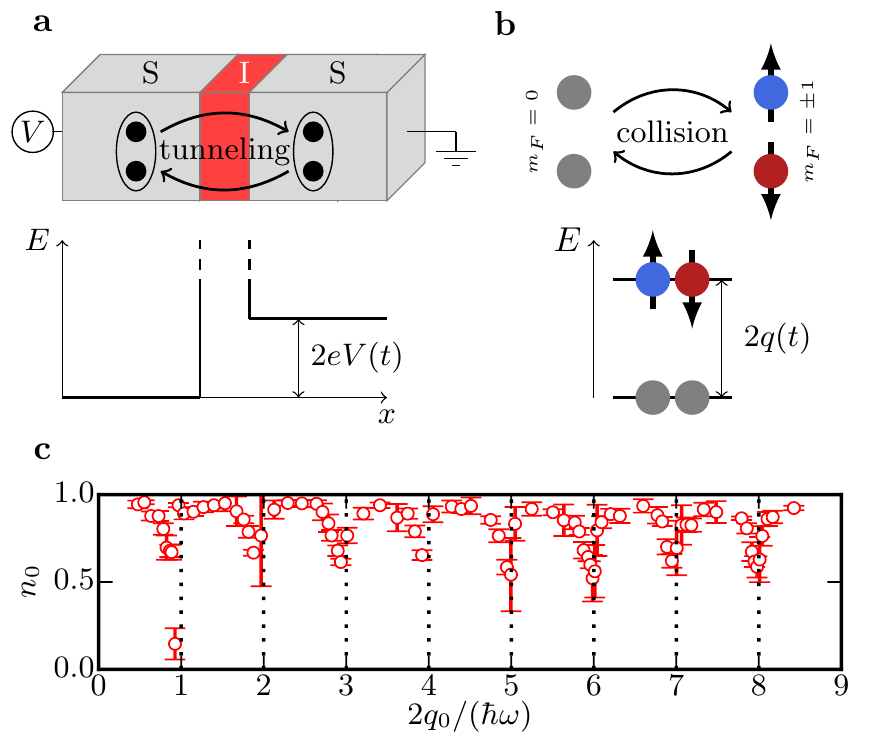}
	\caption{Analogy between two physical systems exhibiting macroscopic quantum coherence: superconducting Josephson junctions (SCJJs--{\textbf{a}}) and spin-1 atomic Bose-Einstein condensates (BECs--{\textbf{b}}). For SCJJs (respectively, BECs), tunneling through the barrier (resp., spin-mixing interactions) generates an electric current (resp., a spin current) controlled by the relative phase across the barrier (resp., between the Zeeman components of the spin-1 wavefunction). An external energy bias $E(t)$ controls the rate of change of the relative phase : the electrostatic energy $E(t)=2e V(t)$ for SCJJs, with $V$ the voltage and $2e$ the charge of a Cooper pair, or the quadratic Zeeman energy $E(t)=2 q(t)$ of a pair of $m_F=\pm 1$ atoms for spin-1 BECs. If the energy bias is modulated around a static value $E_0$, a Shapiro resonance occurs when the modulation frequency $\omega$ fulfill a resonance condition $k_0 \hbar\omega= E_0$, with $k_0$ a positive integer. \textbf{c}: Observation of several ($k_0=1-8$) Shapiro resonances in spin-1 atomic condensates after a relaxation time of 30\,s. Here, $n_0$ is the reduced population of the $m_F=.$ state, and $q_0$ is the static QZE.} 
		\label{figure1}
	\end{figure}

	%
	%

	
	\section{Spin-mixing oscillations in spin-1 condensates}	
	
\label{sec:spinmixingnodrivin}
	
	\subsection{Experimental setup}
	
	We prepare a quasi-pure condensate of spin-1 sodium atoms in a crossed optical dipole trap with a condensed fraction $\gtrsim 0.9$. The condensate is initially polarized in the $m_F=+1$ state (except in Section\,\ref{sec:hysteresis}) and immersed in a spatially uniform magnetic field $\bm{B}=B \bm{u}$, where $\bm{u}=\bm{B}/B$ is taken as quantization axis. Here $m_F$ refers to the eigenvalue of $\hat{\bm{f}} \cdot \bm{u}$ with $\hat{\bm{f}}$ the spin-1 angular momentum operator. In this work, we mostly consider quasi-static variations of $\bm{B}(t)$, slow compared to the inverse of the Larmor frequency $\omega_L=\mu_B B/(2\hbar)$, with $\mu_B$ the Bohr magneton. At the single particle level, the atomic spin states then follow adiabatically the changes of $\bm{u}(t)$: If interactions were negligible, an atom prepared in the $m_F$ state at $t=0$ would remain in that state at all times. Our main observable are the relative populations $n_{m_F}$ of the Zeeman sublevels $m_F=0,\pm 1$. We measure these populations using absorption imaging after a time-of-flight in a magnetic field gradient separating the different Zeeman components (``Stern-Gerlach imaging''). The experimental setup, preparation steps and Stern-Gerlach imaging have been described in details in previous publications\,\cite{jacob2012a,zibold2016a}.

The longitudinal magnetization $M_{||}=\langle \hat{\bm{F}} \cdot \bm{u} \rangle$, with $\hat{\bm{F}}$ the total spin operator, is a conserved quantity in spinor gases\,\cite{law1998a,chang2005a,stamperkurn2013a} (assuming that dipole-dipole interactions\,\cite{lahaye2009a} are negligible). In this work, we focus on the case $M_{||} = 0$. Owing to the adiabaticity condition and to the conservation of magnetization \cite{stamperkurn2013a}, the quadratic Zeeman energy (QZE) is (see Appendix\,\ref{app: Zeeman} for more details):
\begin{align}\label{eq:HZeeman}
	\hat{H}_{\textrm{Zeeman}} &= - q \hat{N}_0,
\end{align}
with $q=\alpha_q B^2$ and $\alpha_q\approx h \times 277\,$Hz/G$^2$. 
		
In the experiments described in the following, we initiate spin mixing dynamics by rotating the internal state of the spin-polarized BEC. This spin rotation is the only exception to the adiabaticity condition indicated above. Experimentally, we apply a radiofrequency field resonant at the Larmor frequency for a time $t_{\pi/2} \approx40\,\mu$s, resulting in a rotation by an angle of $\pi/2$ around an axis orthogonal to the quantization axis $\bm{u}$. With the Zeeman state $m_F=+1$ as starting point, the internal state after rotation is $\frac{1}{2}\left( \vert m_F=+1 \rangle+\vert m_F=-1 \rangle\right)+\frac{1}{\sqrt{2}} \vert m_F=0 \rangle$, hence an initial population $n_{0,\textrm{i}} =1/2$ and $M_{||}=0$.

\subsection{Spin-1 Bose-Einstein condensates in the single-mode regime}

Our experiments are performed in the so-called \textit{single-mode regime} of spinor condensates\,\cite{law1998a,yi2002a,barnett2010a}. This regime is realized for tight traps, such that the lowest energy states correspond to different spin states but the same single-mode spatial orbital (see Appendix\,\ref{app: interactions}). In the single-mode limit and for $M_{||}=0$, the mean-field spin energy per atom is
\begin{align}\label{eq:Espin}
	E_{\textrm{spin}}=-q(t) n_0+ U_s n_0(1-n_0)(1+\cos\theta).
\end{align}
The energy $E_{\textrm{spin}}$ depends only on the relative phase 
\begin{align}
	\theta &  = \phi_{+1}+\phi_{-1}-2 \phi_0,
\end{align}
and on the reduced population $n_0$. The rate of change $\hbar \dot{\theta}$ can be interpreted as a chemical potential difference driving the ``reaction'' $(m_F=+1)+(m_F=-1) \leftrightarrow 2\times  (m_F=0) $, with a ``chemical equilibrium'' reached for $\theta=0$ or $\pi$. For a static QZE $q_0 >0$ and antiferromagnetic interactions $U_s>0$, the classical Hamiltonian $H_{\textrm{spin}}$ is minimal for the so-called polar state with $n_0=1$ that minimizes separately the Zeeman and interaction terms in Eq.\,(\ref{eq:Espin}). 
	
The mean-field spin energy $E_{\textrm{spin}}$ can be interpreted as a classical Hamiltonian for the two conjugate dynamical variables $n_0$ and $\phi$. The equations of motion describing a spin-1 BEC in the single mode limit\,\cite{zhang2005a},
	\begin{align}
	\hbar\dot{n}_{0} &= 2U_s\, n_{0}(1-n_{0})\sin\theta\,,
	\label{eq.SpinMixEq1}\\
	\hbar\dot{\theta} &= -2q(t)+2U_s\,(1-2n_{0})\,(1+\cos\theta)
	\label{eq.SpinMixEq2}
	\end{align}
then correspond to the Hamilton equations of motion, $\hbar\dot{n}_{0}/2 =- \partial E_{\textrm{spin}}/\partial\theta$ and $\hbar\dot{\theta}/2 =\partial E_{\textrm{spin}}/\partial n_0$ \cite{zhang2005a}. 

Eqs.\,(\ref{eq.SpinMixEq1},\ref{eq.SpinMixEq2}) contain the two main ingredients for Josephson physics\,\cite{leggett2001a}. First, the ``spin current'' $\dot{n}_0$ is generated by \textit{coherent} spin-mixing interaction processes controlled by the phase $\theta$. This is analogous to the celebrated Josephson relation $I_s \propto \sin(\Delta\phi)$ linking the supercurrent $I_s$ in a SCJJ to the sine of the relative phase $\Delta\phi$ between the two superconductors on each side of the junction. Second, the external bias $q$ (analogous to the voltage $V$ across the junction) controls the rate of change $\dot{\theta}$ of the relative phase according to Eq.\,(\ref{eq.SpinMixEq2}), analogous to the relation $\hbar\dot{\Delta\phi}=2e V$ with $2e$ the charge of a Cooper pair. 
	
The differences between Eqs.\,(\ref{eq.SpinMixEq1},\ref{eq.SpinMixEq2}) and the ``standard'' Josephson relations reflect some aspects of the physics of atomic gases absent in SCJJs. First, atomic gases can be viewed as closed systems, and Josephson-like phenomena typically lead to population oscillations of large amplitude (comparable to the total atom number), and not to a steady current as for superconducting circuits connected to charge reservoirs. This is reflected in the factor $n_0(1-n_0)$ in Eq.\,(\ref{eq.SpinMixEq1}), which enforces $n_0 \in [0,1]$.  Second, interactions can alter the resonance and the dynamics of the phase, as described by the last term of Eq.\,(\ref{eq.SpinMixEq2}). A similar term plays a major role in double-well realizations of Josephson physics with cold atoms \cite{albiez2005a,levy2007a,leblanc2010a,ryu2013a,valtolina2015a,pigneur2018a}.
	
	\begin{figure}
		\includegraphics[]{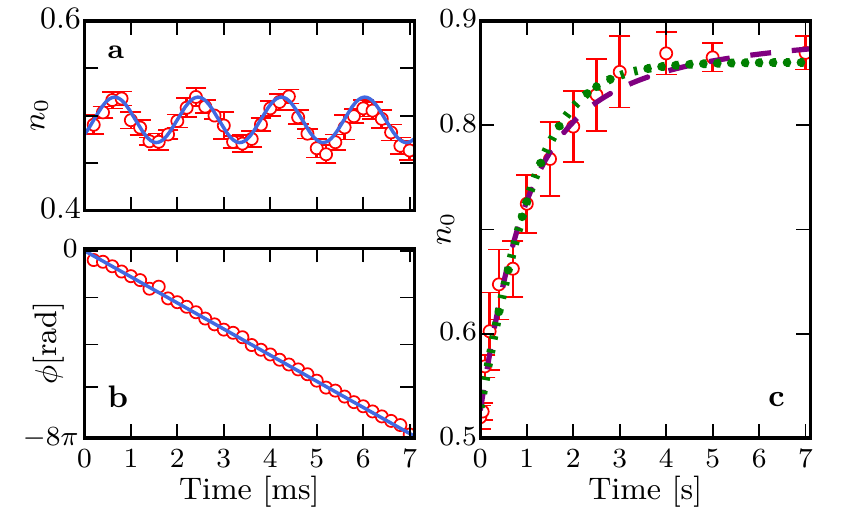}
		\centering
		\caption{\textbf{a-b}: Spin-mixing oscillations without driving in the Zeeman regime $q_0 \gg U_s$. We show the evolution of the population $n_0$ (\textbf{a}) and of the relative phase $\theta$ (\textbf{b.}). \textbf{c}: Relaxation of $n_0$ at long times. We have compared the dissipative model 1 [Eq.\,(\ref{eq.relaxDM1undriven})] and 2 [Eqs.\,(\ref{eq.relaxDM2undriven})] to the experimental data. We take a condensed fraction $f_c <1$ into account using Eq.\,(\ref{eq:n0mes}). The best fits are shown for DM1 (dotted green line), with free parameters $\tau_1=0.18(2)\,$s,  $f_{c,1}\approx 0.85(2)$ and DM2 (dashed purple line) with $\tau_2=0.86(1)\,$s,  $f_{c,2}\approx 0.80(2)$. The fitted damping times correspond to phenomenological damping parameters $\beta_1 \approx 0.20(2)$ and $\beta_2 \approx 1.30(15)\times10^{-3}$.} 
		\label{figure2}
	\end{figure}
	
\subsection{Spin mixing oscillations in the running phase regime and AC Josephson effect}	\label{subsec:spinmixingundriven}

In this paper, we focus on the situation where the static bias $q_0/h\sim 300\,$Hz is much larger than $U_s/h \sim 30\,$Hz. We discuss in this Section the static case without driving, where $q=q_0$ is constant in time. In the regime $q_0 \gg U_s$ (called Zeeman regime in \cite{kronjaeger2005a}), the QZE determines the phase evolution up to small corrections, $\theta(t) \approx \theta(0)-2 q_0t/\hbar$. Eq.\,(\ref{eq.SpinMixEq1}) then predicts harmonic oscillations of $n_0$ at the frequency $\approx 2q_0/\hbar$, with a small amplitude $\propto U_s/q_0$\,\cite{chang2005a,zhang2005a,kronjaeger2005a,kronjaeger2006a,black2007a}. These oscillations constitute the analogue for spinor gases of the AC-Josephson effect: A constant DC bias induces a periodic AC current. 	

We use the method introduced in\,\cite{zibold2016a} to measure the phase $\theta$ and to characterize completely the BEC dynamics. We measure the variance of the transverse spin $\bm{M}_\perp$,
\begin{align}
\langle M_\perp^{2}\rangle=2N^2 n_0(1-n_0)(1+\cos\theta),
\end{align}
from which $\cos\theta$ can be determined. Here $N$ is the total atom number and $\langle \cdot\rangle$ denotes a statistical average over many realizations of the experiment (typically 10-20 in our measurements). This measurement alone cannot determine unambiguously the phase $\theta$. We lift the ambiguity by assuming that $\theta$ wraps monotonously around the unit circle to obtain the curve shown in Fig.\,\ref{figure2}b.

\subsection{Dissipation and relaxation of spin mixing oscillations}
\label{sec:dissipationundriven}
	
	For long times, the spin mixing oscillations are damped and the population $n_0$ eventually relaxes to the expected equilibrium value $n_0 \approx 1$ (see Fig.\,\ref{figure2}c). This relaxation, first observed in \cite{liu2009b}, corresponds to a \textit{loss of energy} of the spinor BEC. Eqs.\,(\ref{eq.SpinMixEq1},\ref{eq.SpinMixEq2}) describe an Hamiltonian dynamics where the energy is a constant of motion \cite{zhang2005a}. As a result, a point or an orbit of the classical phase space $(n_0,\theta)$ cannot be attractive, and relaxation cannot occur within the mean-field framework. 
	
	However, experimental systems are never perfectly isolated, and their coupling to (many) other degrees of freedom playing the role of an energy reservoir enable energy dissipation and thermalization. A plausible candidate to fulfill this role in experiments with ultracold atoms are uncondensed atoms, inevitably present at finite temperatures and forming a bath of collective excitations interacting with themselves and with the condensate. We expect that the interaction of the BEC with this bath acts to restore thermodynamic equilibrium, (\textit{i.e.} a BEC with all atoms in $m_F=0$ for $q_0>0$) with a small decrease of the condensed fraction $f_c$. This is indeed what we observe in Fig.\,\ref{figure2}c, with a typical relaxation time $\sim 1\,$s that depends on $q_0$\,\cite{liu2009b}. 
	
	A first-principle theoretical description of this thermalization dynamics would require to go beyond the Bogoliubov\,\cite{ueda2001a,uchino2010a,ueda2012a} or classical field\,\cite{polkovnikov2011a} descriptions that are only applicable at short times. In this work, we study relaxation over several seconds, \textit{i.e.} several hundred times the intrinsic time scale $\hbar/U_s \sim30\,$ms set by interactions. To the best of our knowledge, no general framework is available to describe strongly out-of -equilibrium dynamics for single-component gases, let alone spin-1 systems. In order to describe the experimental observations and gain some insight on the dynamics, we take in this work a phenomenological approach where a non-conservative term is added ``by hand'' to the Hamiltonian equations of motions \cite{levy2007a,pigneur2018a,liu2009b,kohler2003a}. 
	
	The phenomenological models that we consider in this article generalize the spin mixing Eqs.\,(\ref{eq.SpinMixEq1},\ref{eq.SpinMixEq2}) as follows,
\begin{align}
\label{eq.SpinMixDiss1}
\dot{n}_0 = -\frac{2}{\hbar} \frac{\partial H_{\textrm{spin}}}{\partial \theta} + 	\left.\dot{n}_0\right\vert_{\rm diss},\\
\label{eq.SpinMixDiss2}
\dot{\theta} = \frac{2}{\hbar} \frac{\partial H_{\textrm{spin}}}{\partial n_0} + 	\left.\dot{\theta}\right\vert_{\rm diss}.
\end{align}
The first dissipative model (DM) that we consider was originally proposed in Ref.\,\cite{liu2009b}, 
\begin{align}
\textrm{DM\,1 :}	&\left.\dot{n}_0\right\vert_{\rm diss} = 0, \,\,\, \left.\dot{\theta}_0\right\vert_{\rm diss}=\beta_1 \dot{n}_0.
\label{eq.dissipterm1}
\end{align}
Liu \textit{et al.} correctly argue that the dissipative term in Eq.\,(\ref{eq.dissipterm1}) is the only term linear in $n_0,\theta,\dot{n_0}$ or $\dot{\theta}$ that can explain their measurements\,\cite{liu2009b}. Anticipating on the results in the driven case that will be presented later, we have found that the dissipative model 1 can describe well our experiments without driving, but fails to predict the observed steady state in the strongly driven case. This motivated us to explore other dissipative models, not necessarily linear in $n_0,\theta$ or their derivatives. We propose in this article the alternative
\begin{align}
\textrm{DM\,2 :}	 & \left.\dot{n}_0\right\vert_{\rm diss}=-\beta_2 n_0(1-n_0)\dot{\theta}, \,\,\,
&	\left.\dot{\theta}_0\right\vert_{\rm diss} = 0. \label{eq.dissipterm2}
\end{align}

The dimensionless phenomenological constants $\beta_1,\beta_2$ are real numbers, which are chosen positive to ensure that the energy $H_{\textrm{spin}}$ always decreases (see Appendix\,\ref{app:power}). In the context of cold atoms, formally similar dissipative terms have been proposed previously \cite{kohler2003a,levy2007a,pigneur2018a}, mainly in analogy with terms describing Ohmic dissipation in SCJJs. The DM1 corresponds to a resistor connected in series with the junction, and the DM2 to a resistor in parallel with the junction (``resistively shunted junction model'').
	
Both dissipative models share the same qualitative behavior: they induce a decrease of the total energy (see Appendix\,\ref{app:power}), or equivalently an increase of $n_0$ when $q_0>0$, while conserving the fixed points of the dynamics. For long times (see Appendix\,\ref{app:relaxation_undriven}), the solution of the DM1 is well approximated by
\begin{align}\label{eq.relaxDM1undriven}
\textrm{DM\,1 :}\,\,\,	\overline{n}_0 \approx  1-\frac{\tau_1}{t},
\end{align}
with $\tau_1 = \hbar q_0/(\beta_1 U_s^2)$, while the DM2 predicts (Appendix\,\ref{app:relaxation_undriven})
	\begin{align}\label{eq.relaxDM2undriven}
\textrm{DM\,2 :}\,\,\,	\overline{n}_0=\frac{n_{0,\textrm{i }}}{n_{0,\textrm{i }}+(1-n_{0,\textrm{i }})e^{-t/\tau_2}}.
	\end{align}	
with $\tau_2=2\hbar/(\beta_2  q_0)$. Here $\overline{\cdot}$ denotes a coarse-grained average over a time long compared to a period of the spin-mixing oscillation $\sim \hbar/(2q_0)$, but short compared to the relaxation times $\tau_{1/2}$.

We have compared the predictions of the two models to the experimental results shown in Fig.\,\ref{figure2}c. For this comparison, we account for a small but non-zero thermal fraction. The measured population in $m_F=0$ can be written
\begin{align}\label{eq:n0mes}
n_0 = f_c n_{0,c} +n_0',
\end{align}
with $n_{0,c}=N_{0,\textrm{c}}/N_c$ (resp. $n_0'$) the fraction of condensed (resp. uncondensed) atoms in $m_F=0$. Here $N_{m_F,\textrm{c}}$ denotes the population of condensed atoms in the $m_F$ state, $N_c=\sum_{m_F} N_{m_F,\textrm{c}}$ the number of condensed atoms, $f_c = N_c/N$ the condensed fraction and $N$ the total atom number. We assume for simplicity that thermal atoms are distributed equally among all Zeeman sublevels, so that $n_0' = (1-f_c)/3$. 

We use Eq.\,(\ref{eq:n0mes}) in combination with the dissipative models 1 or 2 for $n_{0,\textrm{c}}$ to fit the experimental data in Fig.\,\ref{figure2}c, using $f_c$ and the relaxation times $\tau_{1/2}$ as free parameters. We find comparable best-fit parameters for both models : $f_c \approx 0.85(2)$, $\tau_1 \approx 0.18(2)\,$s for DM1, 
$f_c \approx 0.80(2)$, $\tau_2 \approx 0.86(10)\,$s for DM2.  The corresponding phenomenological damping parameters are $\beta_1 \approx 0.20(2)$ and $\beta_2 \approx 1.30(15)\times10^{-3}$. The two dissipative models fit well our measurements in Fig.\,\ref{figure2}c, with a small difference more pronounced at long times, but not statistically significant. We conclude that discriminating between the two models is difficult in the undriven case. We will see later in the article that this is no longer the case in the driven case, where the differences are spectacular at long times.


\section{Non-Linear Shapiro Resonances }
\label{sec.ShortTimeDyn}

\subsection{Observation of Shapiro resonances}

\begin{figure}
	\includegraphics[]{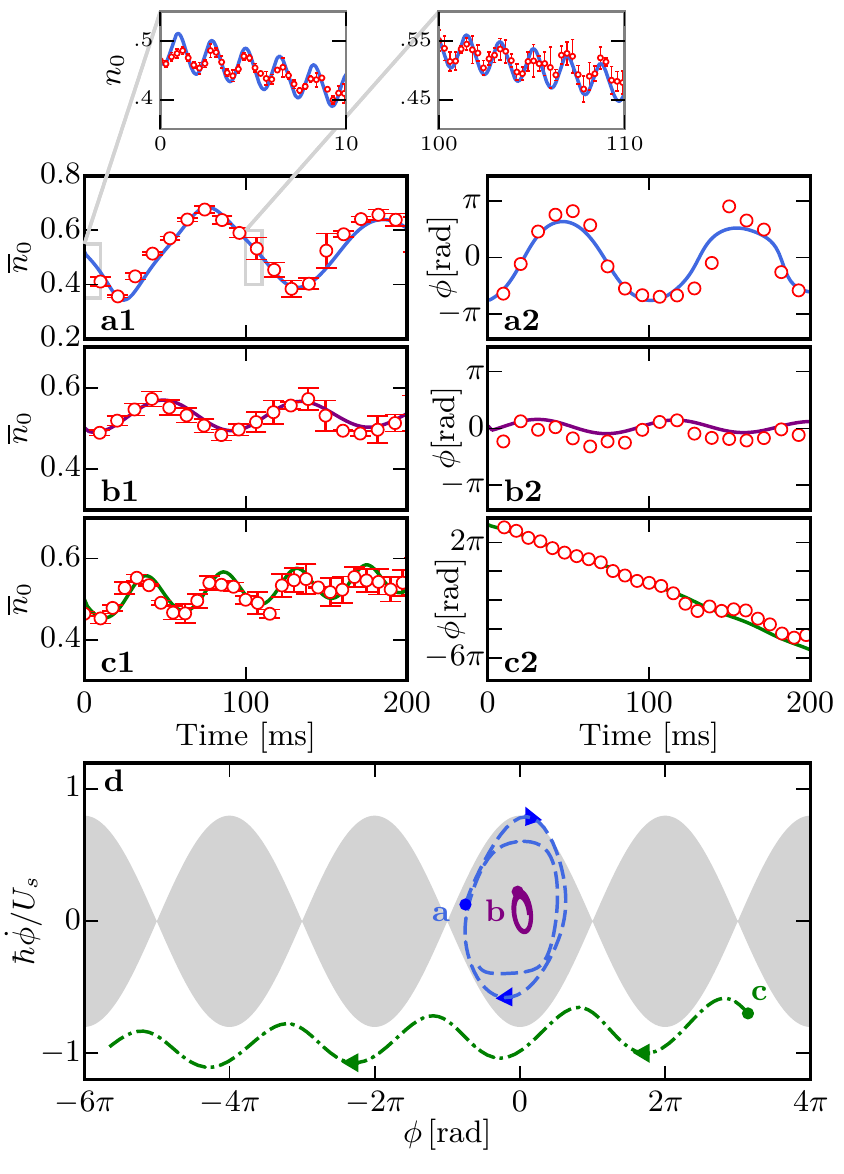}
	\centering
	\caption{Observation of secular oscillations near the first Shapiro resonance $k_0=1$. We show the relative population $\overline{n}_0$ ($\textbf{a-c1}$) and phase $\phi$ ($\textbf{a-c2}$) versus time. For the main panels, the observation times are integer multiples of the modulation period. The curves therefore represent a stroboscopic observation of the dynamics, free of the additional micromotion. The two insets in $\bf{a1}$ (with a smaller time sampling) show the micromotion around the main secular oscillation. The curves in $\textbf{a1-2}$,$\textbf{b1-2}$ correspond to the oscillating regime of the pendulum model, while $\textbf{c1-2}$ corresponds to the clockwise-rotating regime. For all curves, the static bias is $q_0/h=276\,$Hz, the modulation amplitude $\Delta q/h=43.6\,$Hz ($\kappa\simeq 0.08$), and $U_s/h \approx 30\,$Hz. The detuning is $\delta=2\pi\times -5.7\,$Hz ($\textbf{a1-2}$,$\textbf{b1-2}$) and $18\,$Hz ($\textbf{c1-2}$). For curves $\textbf{b1-2}$, we varied the initial phase (see text) to be in the harmonic regime: $\theta(0)=-0.5(2)\,\textrm{rad}$ for $\textbf{a1-2,c1-2}$ and $1.45(2)\,\textrm{rad}$ for $\textbf{b1-2}$. The lines show the numerical solutions of the dissipative model 2 [Eq.\,(\ref{eq.dissipterm2})] with $\beta_2=1.3\cdot10^{-3}$. The calculated curves are further averaged to account for experimental fluctuations (see text). The last panel $\textbf{d}$ shows a phase-space portrait of the trajectories in the $(\phi,\dot{\phi})$ plane, with $\dot{\phi}$ calculated from Eq.(\ref{eq.TimeAvgEq2}). The dashed blue, solid purple and dashed-dotted green line correspond to \textbf{a1-2}, \textbf{b1-2} and \textbf{c1-2}, respectively. The shaded area covers the phase-space area explored in the oscillating regime of the pendulum model.} 
	\label{figure3}
\end{figure}

We now turn to the core of this article where a modulation of the QZE $q(t)$ drives the spinor dynamics. We use a sinusoidal modulation of the QZE around a bias value $q_0$ according to
\begin{align}\label{eq.q}
q(t)&=q_0+\Delta q \sin(\omega t +\varphi_{\rm mod})\Theta(t),
\end{align}
with $\Theta(t)$ the Heaviside step function (see Section\,\ref{sec:phasephi} below). Experimentally, the $x$ component $B_{x}$ of the magnetic field is static, and the $y$ component $B_y=\Delta B \cos[(\omega t+\varphi_{\textrm{mod}})/2+\pi/4]\Theta(t)$ is modulated in a sinusoidal fashion. From Eq.\,(\ref{eq:HZeeman}), the QZE is given by Eq.\,(\ref{eq.q}), with $q_0=\alpha_q (B_{x}^2+\Delta B^2/2)$ and $\Delta q=\alpha_q \Delta B^2/2$.

In a perturbative picture, spin-mixing resonances occur when a pair of atoms in $m_F=0$ can be resonantly transferred to a pair $m_F= \pm 1$ by absorbing an integer number $k$ of modulation quanta, \textit{i.e.} when $k\hbar\omega = 2q_0$ with  $k \in \mathbb{N}$. We define the detuning by
\begin{align}
\hbar\delta = 2q_0-k_0\hbar\omega,
\end{align} 
with $k_0$ the closest integer to $2 q_0/(\hbar\omega)$. 

The left column of Fig.\,\ref{figure3} shows how the population $n_0$ evolves in time for several values of the modulation frequency $\omega$ close to the first resonance with $k_0=1$, such that $\delta\ll q_0$. The dynamics of $n_0$ can be described as a fast (frequency $\simeq 2q_0/\hbar$), small amplitude micromotion, visible in the inset of Fig.\,\ref{figure3}\textbf{a1}, on top of a slow, large amplitude oscillation. The period of the slow oscillation is a hundred milliseconds or more, much longer than the intrinsic timescales set by the QZE or the spin-dependent interactions. This slow dynamics is the result of the coherent build-up over hundreds of periods of the micromotion. The slow ``Shapiro oscillations'' observed near resonance can be viewed as the counterpart for a closed system of the DC current observed near conventional Shapiro resonances in modulated SCJJs.

Fig.\,\ref{figure5} shows the generic behavior observed for longer times, where we observe \textit{(i)} a damping of the contrast of the oscillations on a time scale of several hundred ms, and \textit{(ii)} a drift of the baseline value of $n_0$ towards the equilibrium value without driving, $n_0=1$. We attribute the damping \textit{(i)} mainly to fluctuations of the experimental parameters, leading to shot-to-shot fluctuations of the period and amplitude of the oscillations and therefore to their dephasing after averaging over several realizations of the experiment. We believe the main contribution comes from small ($\Delta N/N\sim 8\,$\%) fluctuations of the atom number. These fluctuations induce fluctuations $\Delta U_s/U_s\sim 6\,$\% of the $N-$dependent interaction strength $U_s$ [see Appendix\,\ref{app.UsVsN} for a calibration of the dependence $U_s(N)$]. We obtain the theoretical curves of Fig.\,\ref{figure3} and Fig.\,\ref{figure5} by solving numerically Eqs.(\ref{eq.SpinMixEq1},\ref{eq.SpinMixEq2}) with the dissipative term (\ref{eq.dissipterm2}) for different $U_s$, and averaging over a Gaussian distribution of $U_s$ with mean and variance deduced from the measured atom number. For short times, this procedure accounts for the observed damping of the oscillations, which occurs on times short enough to neglect the effect of dissipation.

In the remaining of this Section, we first focus on the initial oscillations shown in Fig.\,\ref{figure3}, neglecting the role of dissipation, and postpone the discussion of relaxation at long times \textit{(ii)} to Sec.\,\ref{sec:dissipation}.

\subsection{Secular equations for near-resonant driving}

For our experimental situation with $q_0 \gg U_s$ and for a modulation frequency close to the $k_0$ Shapiro resonance ($\vert \delta \vert \ll q_0$), we derive in Appendix\,\ref{app:secular} effective equations of motion for the slowly evolving components by averaging over the fast micromotion. These secular equations of motion read
\begin{align}
\hbar\dot{\overline{n}_0}&=2 \kappa U_{s}\overline{n}_{0}(1-\overline{n}_{0})\sin\phi,	\label{eq.TimeAvgEq1}\\
\hbar\dot{\phi}&=-\hbar\delta+2U_{s}(1-2\overline{n}_{0})(1+\kappa\cos\phi).	\label{eq.TimeAvgEq2}
\end{align}
Here, $\overline{n}_{0}$ is the time average of $n_0$ over one period $2\pi/\omega$, and $\kappa=J_{k_0}(2\Delta q/\omega)$ is a renormalization factor, with $J_k$ the $k$th-order Bessel function of the first kind. Our modulation scheme is limited to $\Delta q <q_0$. Together with the secular approximation, this implies that $0 < \kappa <1$.
Finally, the secular phase $\phi$ is related to the time-average $\overline{\theta}$ of the phase, 
\begin{align}\label{eq:secularphase}
\phi = \overline{\theta} +k_0(\omega t+\varphi_{\rm mod}+\pi/2).
\end{align}

The secular equations Eqs.\,(\ref{eq.TimeAvgEq1},\ref{eq.TimeAvgEq2}) have a structure similar to the original spin-mixing Eqs.\,(\ref{eq.SpinMixEq1},\ref{eq.SpinMixEq2}) with the replacements $q\to-\hbar\delta/2$ and $e^{i\theta}\to \kappa e^{i\phi}$. Accordingly, Eqs.\,(\ref{eq.TimeAvgEq1},\ref{eq.TimeAvgEq2}) derive from the classical energy of the secular motion,
\begin{align}
E_{\textrm{sec}}=-\frac{\hbar\delta}{2}n_0+U_s n_0(1-n_0)(1+\kappa\cos\phi),
\label{eq:Esec}
\end{align}
using the same Hamilton equations as in the undriven case.

 The different dynamical regimes are best understood in the limit of small driving, $\kappa \ll 1$. We show in Appendix\,\ref{app: Pendulum} that the secular equations Eqs.\,(\ref{eq.TimeAvgEq1},\ref{eq.TimeAvgEq2}) reduce for $\kappa \to 0$ to the ones describing the motion a rigid pendulum of natural frequency $\Omega = \sqrt{2\kappa}U_s/\hbar$, with $\phi$ the angle of the pendulum. The pendulum admits two dynamical regimes, either oscillations around the stable equilibrium point $\phi=0$, or full-swing rotations with $\phi$ running from 0 to $2\pi$. At the transition between the two regimes, the period of the oscillations diverges and the amplitude of the velocity (or, equivalently, of $n_0$) oscillations is divided by two. The same qualitative conclusions hold outside of the weak driving limit, although the position of the separatrix between the two regime is slightly shifted. From Eq.\,(\ref{eq:secularphase}), we note that the regime of small oscillations ($\phi \approx 0$) corresponds to an atomic phase $\overline{\theta} \approx -k_0(\omega t+\varphi_{\rm mod}+\pi/2)$ locked to the drive. Conversely, the regime of full-swing rotations ($\phi \approx - \delta t$) corresponds to a free-running atomic phase $\overline{\theta} \approx -2 q_0 t/\hbar$, barely affected by the drive.

\subsection{Measurement of the secular phase $\phi$}
\label{sec:phasephi}
The two dynamical regimes are best distinguished in the evolution of the phase $\phi$, since the population $\overline{n_0}$ oscillates in both cases. We measure the secular phase using a variant of the method of Section\,\ref{subsec:spinmixingundriven} which allows us to lift the phase ambiguity. We measure $\cos\theta$ as before but for two different times $t_{p}=p\times 2\pi/\omega$ and $t_{p}+T/4=(p+1/4)\times 2\pi/\omega$ with $p \in \mathbb{N}$ and $T=2\pi/\omega$ the period of the modulation. Assuming $\phi(t_{p})\approx \phi(t_{p}+T/4)$ (in accordance with the secular approximation), we obtain, after converting $\theta$ to $\phi$ using the definition of the latter in Eq.\,(\ref{eq:secularphase}), a simultaneous measurement of $\sin\phi(t_p)$ and $\cos\phi(t_p)$ at stroboscopic times $t_p$.

Obtaining confidence intervals on the measurement of $\phi$ is far from obvious. The statistical spread of $\sin\phi(t_p)$ and $\cos\phi(t_p)$ determined by our method increases in time. This can be quantified by computing $S=\langle\cos\phi\rangle^2+\langle\sin\phi\rangle^2$. This quantity is equal to 1 if $\phi$ is perfectly determined and vanishes for $\phi$ completely random in the limit of infinitely large samples. Experimentally we find that $S$ decreases with a characteristic time scale $\sim200\,$ms. We believe the main reason for this decay is the fluctuations of $U_s$ coming from atom number fluctuations. We have computed numerically the probability distribution $\mathcal{P}(\phi)$ of $\phi$ that derives from our expected distribution of $U_s$. Due to the non-linearities of the spin-mixing equation, $\mathcal{P}(\phi)$ broadens rapidly and becomes asymmetric. A Gaussian approximation of $\mathcal{P}(\phi)$ is essentially useless except for very short times (below a few tens of ms). 
We did not pursue a sophisticated statistical analysis accounting for the peculiarities of $\mathcal{P}(\phi)$. Instead we use the quantity $S$ introduced above to estimate when the measurement of the phase becomes unreliable. We restrict the phase measurements to $t \leq 200\,$ms, roughly the interval where $S$ takes values above $1/2$.

In an ideal experiment strictly described by Eq.\,(\ref{eq.q}), the modulation would be turned on instantaneously at $t=0$. The initial phase $\theta(0)=0$ would then be determined by the preparation of the initial state. In practice, a small delay of $\Delta t=100\,\mu$s is present between the preparation and the beginning of the modulation, and the modulation settles to the form in Eq.\,(\ref{eq.q}) after $1-2\,$ms, due to the transient response of the coils used to generate the modulation $B_y$. During this short transient ($\ll \hbar/ U_s$), the populations barely evolve but the phase changes because of the QZE. Both effects can be taken into account in an initial phase shift $\theta(\Delta t)=-(2/\hbar)\times[q_0\Delta t+\int_{0}^{+\infty}[\tilde{q}(t)-q(t)]dt]$, with $\tilde{q}$ the instantaneous QZE actually experienced by the atoms and $q(t)$ the ideal profile. The extra phase shift corresponds to an initial phase $\theta(0) \approx -0.5\,$rad for the data in Fig.\,\ref{figure3}\textbf{a1-2}. We can also insert on purpose a variable delay between the preparation step and the start of the modulation to tune the initial phase. We used this technique to change the initial phase for the data in Fig.\,\ref{figure3}\textbf{b1-2}, which are otherwise obtained for identical conditions as in Fig.\,\ref{figure3}\textbf{a1-2}. 

We plot in Fig.\,\ref{figure3} (right column) the results for $\phi$ for the first resonance $k_0=1$. For small detuning, the phase oscillates around $\phi=0$, \textit{i.e.} the dynamics of the BEC phase is phase-locked with the drive (panels {\bf a1-2,b1-2.}). The excursion of the phase away from $\phi=0$ depends on the detuning and the initial phase, that we can tune (panels {\bf b1-2}) to have $\phi(t=0)\simeq0$. For a given initial phase, when $\delta$ exceeds a critical value corresponding to the transition betweeen the two dynamical regimes, phase locking no longer occurs and the BEC phase runs freely from $0$ to $2\pi$, corresponding to the \textquotedblleft rotating pendulum\textquotedblright  case (panels {\bf c1-2}).

\begin{figure}
	\includegraphics[]{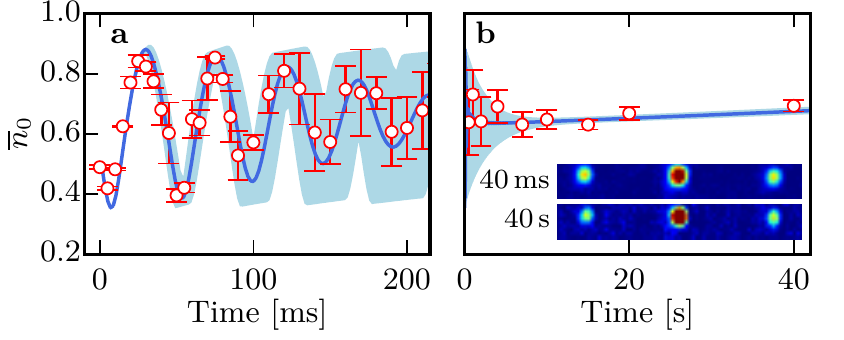}
	\centering
	\caption{\textbf{a}: Damping of Shapiro oscillations. The solid blue curve is calculated from the dissipative model 2 (DM2) and averaged over the fluctuations of $U_s$ caused by atom number fluctuations (see text). The shaded area corresponds to the standard deviation of the distribution of $n_0$ induced by these initial fluctuations.  The static bias is $q_0/h=276\,$Hz, the detuning $\delta=2\pi\times -18\,$Hz, and the modulation amplitude $\Delta q/h=218\,$Hz ($\kappa\simeq 0.36$). The interaction strength is $U_s/h \approx 32\,$Hz for $t=0$ and decays to $\approx 20\,$Hz for $t=40\,$s due to atom losses during the hold time in the optical trap. \textbf{b}: Long-times relaxation of the secular population $\overline{n}_0$ to a steady state. We attribute the small drift of the steady-state population to the decay of $U_s$. 
} 
	\label{figure5}
\end{figure}

\subsection{Period and amplitude of the secular oscillations}

\begin{figure}
	\includegraphics[width=\columnwidth]{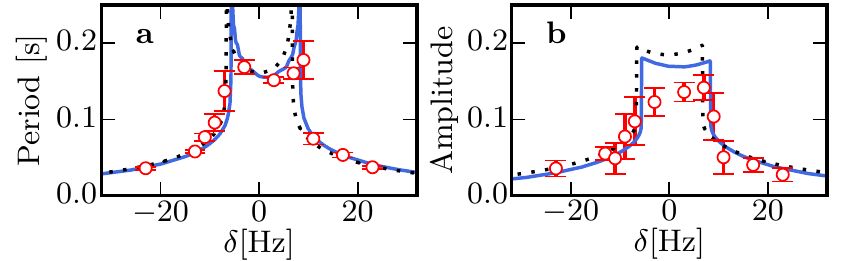} 
	\centering
	\caption{Period ($\bf{a}$) and amplitude ($\bf{b}$) of the secular oscillations versus detuning $\delta$ for the same parameters as in Fig.\,\ref{figure3}. The solid blue lines show the numerical solutions of Eqs.\,(\ref{eq.SpinMixEq1},\ref{eq.SpinMixEq2}), and the dotted black lines the analytical solution of the pendulum model.}	\label{figure4}
\end{figure}

We extract the amplitude and period of the secular oscillations by fitting a periodic function $n_0(t) = \sum_{j=0}^3 a_j \cos(j t/T+\phi_0)$ to the data. We restrict the fit to the first two periods, with the harmonics amplitude $a_j \in \bm{R}$ and the initial phase $\phi_0$ as free parameters. Fig.\,\ref{figure4} shows the period $T$ and amplitude for the first resonance $k_0=1$ versus detuning. The results agree well with a numerical solution of Eqs.\,(\ref{eq.SpinMixEq1},\ref{eq.SpinMixEq2}) (\textit{i.e.}, without taking dissipation into account), and with the pendulum model. Close to resonance, the measured amplitude is systematically lower than the theoretical prediction. This can be qualitatively explained by the presence of uncondensed atoms that do not participate in the coherent secular dynamics.

\section{Long-Time Relaxation and Steady-State}
\label{sec:dissipation}

\subsection{Observation of a Non-Equilibrium Steady State}

\begin{figure}
	\centering
	\includegraphics[]{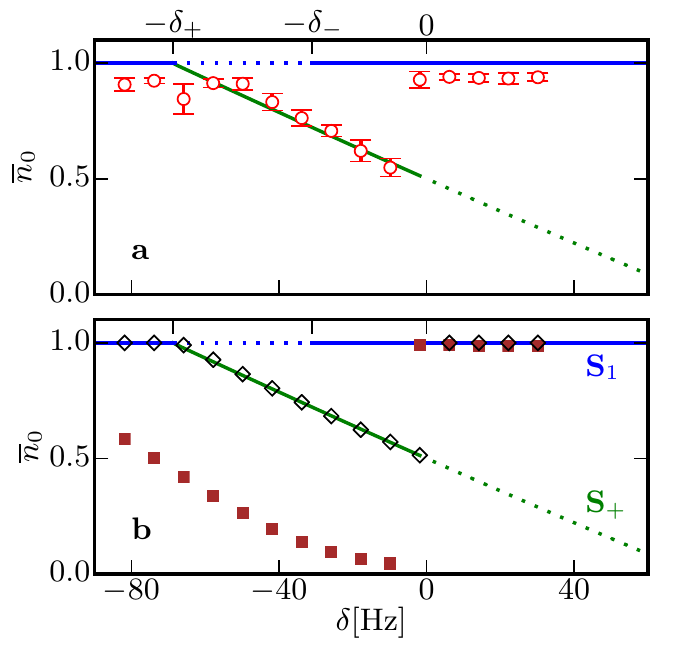}%
	\caption{ {\bf a}: Measured population $n_0$ as a function of detuning $\delta$ after a relaxation time of $10\,$s. The experiment is performed near the first resonance $k_0=1$ ($\omega \approx 2 q_0/\hbar$) and with $n_{0,\textrm{i}}=0.5$. The static bias is $q_0/h \approx 277\,$Hz, the modulation amplitude is $\Delta q/h \approx 227$\,Hz ($\kappa\simeq 0.4$), and the interaction strength is $U_s/h \approx 26$Hz. {\bf b}: Numerical solutions of the dissipative models 1 (Eq.\,\ref{eq.dissipterm1}, brown squares) and 2 (Eq.\,\ref{eq.dissipterm2}, black empty diamonds). In both panels, the horizontal blue (respectively oblique green) line correspond to the stationary state $S_1$ (resp., $S_+$). The solid (resp. dotted) segments correspond to the stability (resp. instability) region according to DM 2 (see Section\,\ref{sec:fixedpoints}). } 

	\label{figure6}
\end{figure}

We now focus on the state reached for long (tens of seconds) evolution times after relaxation has taken place. We find that after the damping of the slow, large amplitude Shapiro oscillations, the population $\bar{n}_0$ reaches a steady state that persists for tens of seconds\,\footnote{The steady state population slightly changes over time due to atom losses and/or evaporation of thermal atoms, that change the condensed atom number and thereby $U_s$ (see Fig.\,\ref{figure5}). The time scale for these changes is very slow (around 10\,s) and modifies significantly $U_s$ from its initial value only after very long times (by about $17$\,\% in 10\,s), much longer than the typical time-scales for the dynamics. We therefore discard these changes for the discussion in the main text.}. Fig.\,\ref{figure6} shows a typical measurement for strong driving ($\kappa=0.38$) near the first resonance $k_0=1$, where we monitor how the steady state value changes as a function of detuning $\delta$. We find that the system relaxes to $\overline{n}_0\approx 1$, except in a range of negative detunings close to the resonance where the population $\overline{n}_0$ takes values between $\approx 0.5$ and $1$. The steady state reached in this strongly driven situation does not correspond to the thermodynamic equilibrium point in the absence of modulation (the ground state of $H_{\textrm{spin}}$ with $\overline{n}_0=1$), nor to the minimum of the secular energy given in Eq.\,(\ref{eq:Esec}) ($\overline{n}_0=1$ for $\delta>0$, $\overline{n}_0=0$ for $\delta<0$). This contrasts strongly with the undriven case where the thermodynamic equilibrium state $n_0\approx 1$ is always observed at long times. 

In the experiments shown in Fig.\,\ref{figure1}\textbf{c}, we show that the same behavior is observed for higher resonances up to $k_0=8$ (limited by the maximal magnetic field we can produce). We set $\omega=2\pi\times100\,$Hz and scan $q_0$ and $\Delta q$ simultaneously keeping $\Delta q/q_0$ (and therefore $\kappa$) approximately constants. After a wait time of $30\,$s, we observe that the system relaxes for all $k_0$ to the same stationary state as for the first resonance. In the following, we therefore concentrate on the case $k_0=1$ as in the previous Section. 


We use the same dissipative models introduced in Section\,\ref{sec:dissipationundriven} to explain the experimental observations. We show in Fig.\,\ref{figure6}\,\textbf{b}
the result of a direct numerical solution (without making a secular approximation) of Eqs.\,(\ref{eq.SpinMixDiss1},\ref{eq.SpinMixDiss2}) for the dissipative models 1 or 2. We observe that the DM 1 fails to reproduce the measured steady-state populations, while the DM 2 predicts a long-time behavior consistent with the experimental results. This contrasts with the undriven case, where both models lead to very similar predictions. In the following, we specialize to the DM 2 and explore its consequences for the long-times steady-state.

\subsection{Fixed points and their stability}
\label{sec:fixedpoints}

\begin{figure}
	\includegraphics[width=\columnwidth]{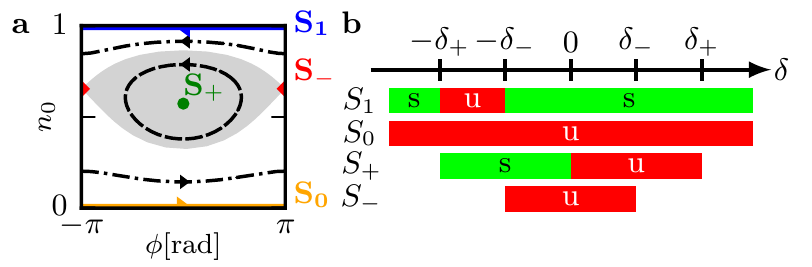} 
	\caption{ Fixed points of the dissipative spin-mixing model 2. \textbf{a}: Phase space portrait of the stationary solutions of Eqs.\,(\ref{eq.TimeAvgEq1},\ref{eq.TimeAvgEq2}). The two limit cycles are labeled $S_0$ ($\overline{n}_0=0$, solid orange line) and $S_1$ ($\overline{n}_0=1$, solid blue line) and the two fixed points $S_+$ (green dot) and $S_-$ (red diamond). The black lines show typical trajectories in the oscillating (dashed line) or rotating (dash-dotted lines) regimes. The shaded area covers the oscillating regime. The plot is shown for $\delta = -2\pi\times10\,$Hz, $U_s=h\times25\,$Hz, $\kappa \simeq 0.38$ ($\delta_-\simeq2\pi\times32\,$Hz) and a damping coefficient $\beta_2 \to 0^+$. \textbf{b}: Table summarizing for $\beta_2 \to 0^+$ the ranges of detuning where each stationary solution is stable ('s') or unstable ('u'). The boundaries $\delta_\pm$ are defined after Eq.\,(\ref{eq:n0pm}).}
		\label{figure7}
\end{figure}

We look for (possibly metastable) secular solutions of dissipative model 2 where the population $\overline{n}_0$ is stationary. We derive generalized secular equations as in Section\,\ref{sec.ShortTimeDyn} starting from Eqs\,(\ref{eq.SpinMixDiss1},\ref{eq.SpinMixDiss2},\ref{eq.dissipterm2}) defining the DM 2. Observing from Eq.\,(\ref{eq:secularphase}) that $\dot{\overline{\theta}} \approx -\omega + \dot{\phi}$, we find
\begin{align}
\hbar\dot{\overline{n}}_0&=\overline{n}_{0}(1-\overline{n}_{0})\left( 2\kappa U_s\sin\phi+\beta_2 \hbar\omega-\beta_2\hbar\dot{\phi}\right).	\label{eq.TimeAvgDissEq1}
\end{align}
The phase dynamics is still determined by Eq.\,(\ref{eq.TimeAvgEq2}). From Eq.\,(\ref{eq.TimeAvgDissEq1}), we identify four possible states for which $\dot{\overline{n}}_0=0$.


The first two states correspond $\overline{n}_0=0, 1$. In these two limiting cases, the relative phase $\theta$ (and thus $\phi$) is physically irrelevant and can take any value. These two solutions labeled $S_{0},S_1$ in the following, correspond to ``limit cycles'' in the language of dynamical systems. The other two stationary states, labeled $S_\pm$, correspond to fixed points of the dissipative equations of motion where $\dot{\overline{n}_0} = \dot{\phi}=0$. They correspond to the secular phases $\phi_+ = \epsilon, \phi_- =  \pi-\epsilon$, where the angle $\epsilon$ obeys $\sin \epsilon = - \beta_2 \hbar\omega/(2\kappa U_s)$. The populations at the fixed points are
\begin{align}\label{eq:n0pm}
\overline{n}_{0,\pm}= \frac{1}{2}\left(1-\frac{\delta}{\delta_\pm}\right),
\end{align}
with $\hbar\delta_{\pm} = 2 U_s (1 \pm \kappa \cos \epsilon)$. Fig.\,\ref{figure7}a shows the location of the stationary solutions in a secular phase-space portrait ($\overline{n}_0,\phi$). One of the two limit cycles correspond to the minima of the secular energy $E_{\textrm{sec}}$ depending on the sign of $\delta$. The fixed point $S_+$ is always the maximum of $E_{\textrm{sec}}$, while $S_-$ is a saddle point.

Dissipation must be present but not too strong to ensure the existence of a fixed point of the dynamics. Indeed, the fixed points $S_\pm$ disappear when $ \beta_2 \geq 2 \kappa U_s/(\hbar\omega)$ ($\sin\epsilon\leq -1$). If $\beta_2$ is too large or $\kappa$ too small, the drive cannot provide enough energy to overcome the dissipation and create a metastable state (see discussion below). This is consistent with other experiments we performed with a weaker driving strength  $\kappa \sim 0.08$, where we found that the relaxation to the fixed point was less robust than the one shown in Fig.\,\ref{figure6}. 

For strong driving (large modulation amplitude $\Delta q\sim q_0$) as in the experiments shown in Fig.\,\ref{figure6}, we find $\phi_+ \approx 0.04$ corresponding to the limit $\epsilon \propto \beta_2 \to 0^+$. In this situation, the positions of the fixed points are well approximated by setting $\cos\epsilon \approx 1$ and $\hbar\delta_{\pm} = 2 U_s (1 \pm \kappa)$. The positions of the fixed points are therefore independent of the precise value of $\beta_2$ to first order in the small parameter $\epsilon$.

We study the dynamical stability of the stationary solutions in App.\,\ref{app:stabilityfixed} for a phenomenological damping coefficient $\beta_2 \to 0^+$. We summarize the results in Fig.\,\ref{figure7}b. The drive destabilizes $S_{1}$ in a small region of positive detunings around the resonance, while $S_0$ is always unstable because of the dissipation. The fixed point $S_+$ is stable only for $\delta<0$, while $S_-$ is always unstable.

At first glance one may expect that energy dissipation induce relaxation to an energy minimum. In fact, the work delivered by the drive compensates for the dissipated energy, thereby stabilizing the system in a highly excited state (App.\,\ref{app:energybalance}). At the fixed point $S_{\pm}$, the atomic phase locks to the drive with a small phase lag such that the power absorbed from the drive exactly compensates the power loss due to dissipation. This phase-locking enabled by dissipation is reminiscent of the dissipative phenomenon leading to Shapiro steps in SCJJS\,\cite{kautz1996a}.

\subsection{Interpretation of experimental results}

We can now interpret the experimental findings of Fig.\,\ref{figure6}. The position of the stable fixed point $S_+$ in the limit $\beta_2 \to 0$ is shown in Fig.\,\ref{figure6}, and explains well the observed steady-state populations for $\delta\in[-\delta_+,0]$. Outside this window, the system relaxes to the equilibrium state $S_1$. We interpret the observed ``trapping'' in the state $S_+$ as follows. A system prepared with $n_{0,\textrm{i}} \approx 0.5$ tends to relax to the ground state $S_1$ of $H_{\textrm{spin}}$, as observed for $\vert \delta \vert > \delta_+$ where there is no fixed point. For $\delta\in[-\delta_+,0]$, the derivative of the phase $\dot{\phi}$ diminishes in absolute value as $\overline{n_0}$ increases because of the dissipation, and it progressively vanishes. At this point, that corresponds to $S_+$, $\dot{\overline{n_0}}$ also vanishes and the system remains trapped in this state.
On the contrary, for $\delta\in [0,\delta_+]$, $S_+$ corresponds to $\overline{n}_{0,+} \leq  1/2$ and $\vert \dot{\phi}\vert$ increases as $\overline{n}_0$ increases. The trajectory tends to move the system away from $S_+$. As a result dissipation acts in this case as in the undriven case, and the system eventually reaches $S_1$ after a sufficient relaxation time. 



The scenario described above explains all observations but one. In Fig.\,\ref{figure1}\textbf{c}, for very small but negative $\delta$ near the first resonance, the system relaxes to $\overline{n}_0\simeq0.16$. This observation is consistent with thermalization in the secular Hamiltonian where the lowest energy state is $\overline{n}_0 \approx 0$ when $\delta \leq 0$ (a non-zero thermal fraction or an incomplete thermalization could explain the deviation from $\overline{n}_0=0$). 


\section{Hysteretic Behavior}
\label{sec:hysteresis}

\begin{figure}
	\includegraphics[]{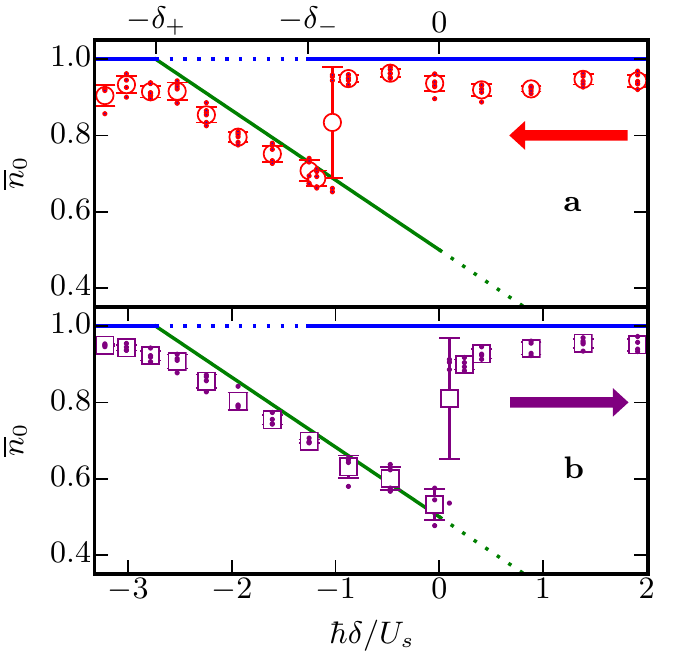} 
	\centering
	\caption{ Observation of hysteresis in the relative population $n_0$ after a detuning ramp. We prepare a spinor BEC with $n_{0,\textrm{i}} \simeq 1$, and scan the detuning by changing $q_0$ for fixed $\omega=2\pi\times277\,$Hz and $\Delta q/h = 227\,$Hz. In \textbf{a} (respectively, \textbf{b}), the ramp decreases (resp., increases) from $\delta_i\approx  2.0 U_s/\hbar $ (resp., $\delta_i\approx-3.3U_s/\hbar$). The horizontal blue (resp., oblique green) line correspond to $S_1$ (resp., $S_+$). The solid (resp., dotted) segments correspond to the stability (resp., instability) regions.}
	\label{figure8}
\end{figure}

According to the stability diagram of Fig.\,\ref{figure7}\textbf{b}, there is no stationary solution that would be stable for all detunings $\delta$. Furthermore, there are two stable solutions $S_+$ and $S_1$ in the interval $[-\delta_-,0]$. In such a situation, one can expect hysteretic behaviour, that we searched for using a slighly different procedure than in the rest of the article. 

We prepared a BEC in the state $m_F=0$, such that $\overline{n}_{0,i}\sim1$ (up to thermal atoms in $m_F=\pm1$). We apply the modulation as before but slowly ramp the static bias $q_0$ over a ramp time of 3\,s, and then hold the driven system at the final $q_0$ value for 7\,s. This amounts to a slow ramp of the detuning $\delta$ decreasing (respectively, increasing) from $\delta_i$ to $\delta_f$ in Fig.\,\ref{figure8}{\bf a} (resp., Fig.\,\ref{figure8}{\bf b}). For decreasing ramps with $\delta_i>\delta_+$, the system remains in $S_1$ in the domain $\delta > -\delta_-$ where $S_1$ is stable. Continuing the ramp further, $S_1$ becomes unstable and we find that the system relaxes to $S_+$ as in the previous experiments. Conversely, for an increasing ramp starting from $\delta_i<-\delta_+$, the system follows $S_+$ while it is stable, $i.e.$ for $\delta_f \in [-\delta_+, 0]$ and $S_1$ otherwise. We therefore observe an hysteresis cycle spanning the interval $\delta\in[-\delta_-,0]$ where both $S_1$ and $S_+$ are stable. 


\section{Conclusion}
\label{sec:conclusion}

In conclusion, we have observed the analogue for a driven spin-1 BEC of the Shapiro resonances characteristic of the AC Josephson effect in SCJJs. The population dynamics near each resonance corresponds to a slow and non-linear secular oscillation on top of a rapid micromotion. We have found that the driven spin-1 BEC relaxes at long times to asymptotic states phase-locked to the drive and that are not stable without it. We proposed a phenomenological model of dissipation that describes quantitatively the relaxation process and its outcome. The dynamics in the driven case allows us to discriminate between different phenomenological models, in contrast to the situation without driving where different models lead to very similar behavior. 

The microscopic origin of the dissipation remains to be investigated. While dissipation probably comes from interaction between condensed and uncondensed atoms, a quantitative description of their interactions and of the resulting thermalization process is lacking. The procedure we used in this paper led to a set of dissipative equations which are essentially generalized Gross-Pitaevskii equations. While we have found excellent agreement between the experimental results and the predictions of these equations, the procedure is purely phenomenological and whether these generalized Gross-Pitaevskii equations can be derived from first principles or not remains an open question. A detailed study could also be useful to understand other types of driven quantum gases where an optical lattice potential\,\cite{eckardt2017a} or the interaction strength\,\cite{clark2017a} are modulated.

Another interesting question is related to the occurence of deterministic chaos in a driven spin-1 BEC\,\cite{cheng2010a}. Without driving, chaotic behavior can be ruled out for spin-1 BECs on the basis of the Poincar{\'e}-Bendixson theorem\,\cite{strogatz_chaos}: The theorem excludes chaotic solutions for a two-dimensional parameter space $(n_0,\theta)$. However, chaos can appear in higher spin $F \geq 2$ systems, as studied in \cite{kronjaeger2008a}, or in driven systems\,\cite{cheng2010a}, where time plays the role of a third variable. When the secular approximation holds, the system is effectively two-dimensional. One thus expects to find chaos in situations where the secular approximation breaks down. Using the non-dissipative spin-mixing equations and adapting the methods of \cite{cheng2010a} to our system, we have found numerically that chaos can be present in the vicinity of Shapiro resonances for strong modulation and small bias, $\Delta q \sim q_0 \sim U_s$. For almost all experiments reported in this paper, where $q_0\gg U_s$, we did not find any evidence of chaotic behaviour. The only exception is the curve in Fig.\,\ref{figure1}$\bf{c.}$, where $q_0 \simeq h\times100\,$ Hz is only three times larger than $U_s$. The deviation from the fixed point near $\delta=0$ for the first resonance could be connected to the onset of chaotic behavior, which is an interesting direction to explore in future work.  

Finally, an interesting application of the driving could be to control dynamically the strength of spin mixing interactions. Spin-mixing interactions work as parametric amplifiers in quantum optics. Such parametric amplifiers are phase-sensitive, and are also known to generate squeezing (see \cite{gross2011a,hamley2012a,luo2017a} for the spinor case) which enable interferometric measurements below the standard quantum limit\,\cite{luecke2011a,linnemann2016a,zou2018a}. A promising direction for the development of interferometers operating at the Heisenberg-limit are the so-called $SU(1,1)$ interferometers\,\cite{linnemann2016a}, which can be viewed as Mach-Zehnder interferometers where the beam splitters are replaced by parametric amplifiers. As shown in Appendix\,\ref{app:quantum}, the quantum version of the secular single-mode Hamiltonian\,\cite{law1998a} is renormalized by driving as in the mean-field framework. This implies that spin-mixing collisions can be enabled by moving close to a Shapiro resonance for a controllable duration, and then disabled by detuning the system away from resonance. Such dynamical control over the spin-mixing process could improve the performances of $SU(1,1)$ interferometers\,\cite{linnemann2016a}.

\begin{acknowledgments}	
\noindent  We would like to thank \c C. Girit, D. Delande, Y. Castin, A. Sinatra, A. Buchleitner and L. Carr for insightful discussions. This work has been supported by ERC (Synergy Grant UQUAM). K.\,J.\,G. acknowledges funding from the European Union's Horizon 2020 Research and Innovation Programme under the Marie Sklodowska-Curie Grant Agreement No. 701894. LKB is a member of the SIRTEQ network of R\'egion Ile-de-France.\end{acknowledgments}

\appendix
\section{Theoretical description of spin-1 condensates}
\subsection{Zeeman energy and adiabatic following}
\label{app: Zeeman}

We consider a gas of spin-1 atoms in a magnetic field $\bm{B}=B(t) \bm{u}(t)$ with time-dependent amplitude $B$ and orientation $\bm{u}$. We take the instantaneous direction $\bm{u}(t)$ of $\bm{B}$ as quantization axis. The label $m_F=0,\pm1$ then corresponds to the instantaneous Zeeman state $\vert m_F \rangle_{\bm{u}}$, \textit{i.e.} the eigenstate of $\hat{\bm{f}}\cdot\bm{u}$ with eigenvalue $m_F$, with $\hat{f}_{x,y,z}$ the spin-1 matrices. The atomic spins precess around $\bm{u}$ at the characteristic Larmor frequency $\omega_L=\mu_B B/2$. The atom internal state follows adiabatically changes of $B$ and $\bm{u}$ slow compared to the inverse Larmor frequency $\omega_L^{-1}$ provided that the adiabatic condition $\dot{\omega}_L \ll \omega_L^2$, where the dot denotes a time derivative, is fulfilled at all times. In our experiment, this condition can also be written $\omega B_y \ll \omega_L \vert \bm{B} \vert$. In most of this work, the Larmor frequency is around $\omega_L \sim 2\pi \times 0.7\,$MHz. Since $B_y \leq \vert \bm{B}\vert$, the sufficient condition $\omega/\omega_L\sim 10^{-3}$ is always fulfilled.

The adiabatic Zeeman energy in a frame aligned with the instantaneous magnetic field is $\hat{h}'_{Z} = p(t) \hat{f}_z + q(t) \left[\hat{f}_z^2 -1\right]$, with $p(t)= -\mu_B B(t)/2$ and $q(t)=\alpha_q \bm{B}^2$. Taking the conservation of magnetization into account, the Zeeman energy experienced by a system of $N$ atoms is given up to a constant term by Eq.\,(\ref{eq:HZeeman}),
with a QZE proportional to $\hat{N}_0=\hat{a}_{0}^\dagger\hat{a}_{0}=\sum_{m,m'} \langle m \vert \hat{f}_z^2-1  \vert m'\rangle \hat{a}_{m}^\dagger\hat{a}_{m'}$ in second quantized notation. Here and in the following, $\hat{a}_m$ denotes the annihilation operator of a boson in the Zeeman state $m$.


\subsection{Spin-dependent interactions}
\label{app: interactions}
We focus here on the so-called \textit{single-mode regime} appropriate to describe our experimental system \cite{law1998a,yi2002a,barnett2010a}. This regime corresponds to a spinor condensate confined in a tight trap, such that the lowest energy states correspond to atoms with different spin states in the same single-mode spatial orbital $\overline{\phi}({\bf r})$. 
The Hamiltonian for a gas of $N$ condensed spin 1 bosons in the SMA can be written
\begin{align}\label{eq:HSMA}
\hat{H}_{\textrm{spin}} &=\frac{U_s}{2N} \bm{\hat{S}}^2-q \hat{N}_0.
\end{align}
Here $U_s$ is a spin-dependent interaction energy determined by the single-mode orbital, $U_s = (4\pi \hbar^2 N a_s)/m_{\textrm{Na}} \times \int d\bm{r}\, \vert \overline{\phi}(\bm{r}) \vert^4$, with $a_s \approx 0.13\,$nm the spin-dependent scattering length\,\cite{knoop2011a} and $m_{\textrm{Na}}$ the mass of a sodium atom.

At low temperatures and for weak interactions, almost all atoms are expected to condense into the same single-particle vector state $\boldsymbol{\zeta}$, a complex vector. The three complex components $\zeta_m=\sqrt{n_m}e^{i\phi_m}$ of the condensate wavefunction are determined by six real numbers. Accounting for the normalization, for an irrelevant global phase, and for the conservation of magnetization leaves only three real variables, the relative population $n_0$ and two relative phases $\theta$ and $\eta=\phi_{+1}-\phi_{-1}$. The latter describes the Larmor precession, and decouples from the other two variables. The mean spin energy in the state $\boldsymbol{\zeta}$ is given by Eq.\,(\ref{eq:Espin}).

\subsection{Calibration of $U_s$}\label{app.UsVsN}
\begin{figure}
	\includegraphics[]{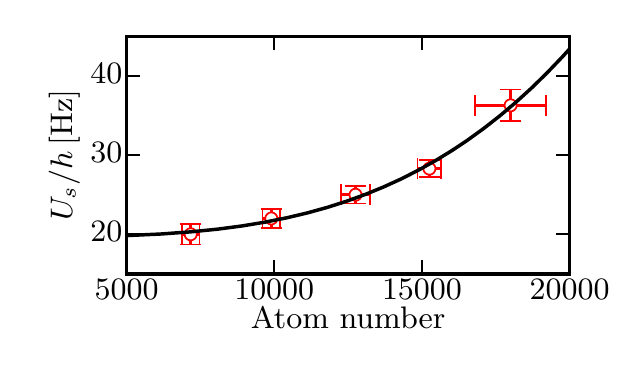} 
	\caption{Interaction strength $U_s$ measured for different atom number. The black solid line is an heuristic fit (see main text).
The QZE is static and equal to $q_0 /h\approx 0.7\,$Hz$\ll U_s$ ($B_x\approx 50\,$mG).	}
	\label{fig.UsVsN}
\end{figure}

We calibrate the interaction strength $U_s$ using the well-established behavior of spin-mixing oscillations without driving. 
For a given total atom number $N$, we fit the observed population oscillations with the numerical solutions of Eqs.\,(\ref{eq.SpinMixEq1},\ref{eq.SpinMixEq2}) treating $U_s$ as a free parameter, all other parameters being kept constant. We shown the fitted value of $U_s$ versus $N$ in Fig.\,\ref{fig.UsVsN}. The dependence on atom number reflect the fact that our experiments are in the crossover between the ideal gas (where $U_s$ is independent of $N$) and the Thomas-Fermi regime (where $U_s \propto N^{2/5}$). We use the heuristic function $U_s(N)/h =  a(1+(N/N_0)^b)$ to calibrate the dependence, with best fit parameters $a\simeq20\,$Hz, $b\simeq3.5$ and $N_0\simeq19\,000$. Small fluctuations of $N$ induce fluctuations of $U_s$ according to $\delta U_s = ab(N/N_0)^b\delta N/\langle N \rangle$. In our experiment, we have typically $\langle N \rangle\simeq 13\,000$ and $\delta N \simeq 1\,000$, which correspond to $\langle U_s \rangle/\hbar\simeq25\,$Hz and $\delta U_s/\hbar \simeq 1.5\,$Hz.

\section{Relaxation of spin oscillations without driving}

\subsection{Calculation of the dissipated power}
\label{app:power}
We compute the dissipated power from $\mathcal{P}_{\rm diss}=\frac{d E_{\rm spin}}{dt} = \frac{\partial E_{\rm spin}}{\partial n_0}\dot{n}_0 + \frac{\partial E_{\rm spin}}{\partial \theta}\dot{\theta}$, or equivalently from
\begin{align}
\mathcal{P}_{\rm diss} &  = \frac{\hbar}{2} \left( \dot{n}_0 \left.\dot{\theta}\right\vert_{\textrm{diss}}-\dot{\theta} \left.\dot{n}_0\right\vert_{\textrm{diss}}\right).
\end{align}
For the dissipative model 1 introduced in the text, the dissipated power $\mathcal{P}^{(1)}_{\rm diss}=-\frac{\hbar}{2}\beta_1 \dot{n}_0^2$ is proportional to the square of the ``current'' $\dot{n}_0$. For the dissipative model 2, the dissipated power is $\mathcal{P}^{(2)}_{\rm diss}=-\frac{\hbar}{2}\beta_2 n_0(1-n_0)\dot{\theta}^2$. In both case it is negative, which corresponds to energy dissipation.

\subsection{Long-time relaxation}
\label{app:relaxation_undriven}

The spin dynamics without driving consists of a ``fast'' evolution of the population and of the relative phase $\theta$ superimposed on a slowly-varying envelope. In the limit $q_0\gg U_s$, the envelope of $n_0$ relaxes to $n_0=1$ over times long compared to the period $\sim \hbar/(2 q_0)$ of spin-mixing oscillations. Averaging in a time window long compared to this period, we obtain effective equations for the envelope of $n_0$ that can be solved analytically. For the dissipative model 1 with the initial condition $n_0(0)=n_{0,\textrm{i}}$, we find that the envelope of $n_0$ obeys 
the implicit equation, $f(\overline{n}_0)=f(n_{0,\textrm{i }})+t/\tau_1$, with $f(x)=2\ln[x/(1-x)]+(2x-1)/[x(x-1)]$ and $\tau_1 = \hbar q_0/(\beta_1 U_s^2)$. For $ t \gg \tau_1$, the solution is well approximated by Eq.\,(\ref{eq.relaxDM1undriven}). For the dissipative model 2, we obtain Eq.\,(\ref{eq.relaxDM2undriven}) by direct integration.

\section{Secular dynamics}
\subsection{Derivation of the secular equations}
\label{app:secular}

In this Section, we derive the secular equations Eqs.\,(\ref{eq.TimeAvgEq1},\ref{eq.TimeAvgEq2}). Integrating formally Eq.\,(\ref{eq.SpinMixEq2}), we rewrite $\theta=\alpha-2p$, where
\begin{align}
p(t)=\frac{1}{\hbar}\int_{0}^{t} q(t')dt'=\overline{p}-\frac{\eta}{2} \cos(\omega t+\varphi_{\textrm{mod}}).
\label{eq.p}
\end{align}
Here $\overline{p}=\frac{q_0 t}{\hbar}+\frac{\chi}{2}$ and $\alpha$ verifies $\hbar\dot{\alpha}=2U_{s}(1-2n_{0})(1+\cos\theta)$. We introduced a modulation index $\eta=2\Delta q/(\hbar\omega)$ and an initial phase $\chi=\eta\cos\varphi_{\textrm{mod}}$. 

We now assume that the driving frequency is close to a parametric resonance, \textit{i.e.} $\omega \sim 2 q_0/(\hbar k_0)$ for some integer $k_0$, and that $q_0\gg U_s$. All physical variables feature in general a large-amplitude secular motion occurring on time scales much longer than the modulation period, plus rapidly-varying terms oscillating at harmonics of $2q_0/\hbar$ that describe the micromotion. In the regime $q_0\gg U_s$, the amplitude $\sim U_s/ q_0$ of the micromotion of $n_0$ and $\alpha$ is small. Taking the time average over one period of the modulation, $\overline{\,\cdot\,}=\frac{1}{T}\int_0^T dt \cdot$, eliminates the micromotion in Eqs.\,(\ref{eq.SpinMixEq1},\ref{eq.SpinMixEq2}),
\begin{align}\label{eq.n0barApp}
\hbar\dot{\overline{n_0}}& \approx 2U_{s} \overline{n_0}(1-\overline{n_0})\overline{\sin \theta},\\
\label{eq.alphaApp}\hbar\dot{\overline{\alpha}}& \approx 2U_{s}(1-2 \overline{n_0})\left(1+\overline{\cos\theta} \right).
\end{align}
We compute the time average of trigonometric functions of $\theta$ using the Jacobi-Anger expansion, $e^{i a \sin(\theta)} = \sum_{k=-\infty}^{+\infty} J_k(a) e^{i k \theta}$, with $J_k$ a Bessel function of the first kind. Neglecting the micromotion of $\alpha$, we can write $\overline{e^{i\theta}}\approx e^{i \overline{\alpha} } \overline{e^{-2ip}}$, with
\begin{align}\label{eq:JA}
e^{-2 i p}&=\sum_{k=-\infty}^{+\infty}J_k(\eta) e^{i \left(-\frac{2q_0}{\hbar}+k \omega\right) t + i k(\phi_{\rm mod}+\pi/2)-i \chi}.
\end{align}
The term $k=k_0$ in the expansion gives rise to a slowly varying secular contribution, while all other terms average out over one period of the modulation. Neglecting the non-resonant terms, we obtain
$\overline{e^{-2 i p}}=\kappa e^{i \zeta(t)}\,,$
with $\hbar\delta = 2q_0-k_0\hbar\omega$, $\zeta(t)= k_0 (\phi_{\textrm{mod}}+\pi/2) -\chi-\delta t$ and $\kappa=J_{k_0}(\eta)$. This finally leads to
\begin{align}\label{eq.eitheta}
\overline{e^{i\theta}}\approx \kappa e^{i\phi}\,
\end{align}
where the secular phase $\phi=\zeta+\overline{\alpha}$ is defined as
\begin{align}
\phi=-\delta t+\overline{\alpha}+k_0( \varphi_{\mathrm{mod}}+\pi/2)-\chi.\label{eq.phi}
\end{align}
Eqs.\,(\ref{eq.TimeAvgEq1},\ref{eq.TimeAvgEq2}) follow from Eqs.\,(\ref{eq.n0barApp},\ref{eq.alphaApp},\ref{eq.eitheta},\ref{eq.phi}).

From Eq.(\ref{eq.phi}), we can relate $\phi$ to the atomic phase, $\overline{\theta}=\phi -k_0(\omega t +\varphi_{\rm mod}+\pi/2)\,$. This equality shows that when $\phi$ is oscillating, $\theta$ also oscillates around the phase of the drive $-k_0(\omega t+\varphi_{\rm mod}+\pi/2)$, up to a constant.

\subsection{Rigid Pendulum Model}\label{app: Pendulum}
In the weak driving regime, $\kappa\ll1$, the $\kappa\cos\phi$ term in Eq.\,(\ref{eq.TimeAvgEq2}) is negligible. Moreover, the amplitude of variation of $\overline{n}_0$ is small. To prove the last point, we integrate Eqs.\,(\ref{eq.TimeAvgEq1},\ref{eq.TimeAvgEq2}) and obtain the implicit equation
$\big[g(x)\big]^{\overline{n}_{0}(t)}_{\overline{n}_{0,i}}=-\kappa\big[\cos x \big]^{\phi(t)}_{\phi_i}$,
with $g(x)=\big(1-\frac{\hbar\delta}{2U_s}\big)\ln\big(\frac{x}{1-x}\big)+2\ln(1-x)\,.$
This implies that the amplitude of variation of $\overline{n}_0$ are indeed small when $\kappa\ll1$. This allow us to linearise Eq.\,(\ref{eq.TimeAvgEq1}). \\
With the initial condition $n_{0,\textrm{i}}=1/2$, we obtain $\hbar\dot{\overline{n}}_0\simeq \frac{\kappa U_s}{2}\sin\phi$. Taking the time derivative of Eq.\,(\ref{eq.TimeAvgEq2}), we then find that the phase obeys the pendulum equation
\begin{align}
\ddot{\phi}+\Omega^2\sin\phi=0\,,
\end{align}
with natural frequency $\Omega=\sqrt{2\kappa} U_s/\hbar$. The angular velocity of the pendulum $\dot{\phi}$ is determined by $\dot{\phi} = -\delta+ 4 U_s(1/2-\overline{n}_0)$.

\subsection{Energy balance}
\label{app:energybalance}
In this Section, we compute the power delivered by the drive in the framework of DM2. In particular, we show that at the fixed points $S_\pm$, it compensates for the dissipated energy.  For simplicity, we focus on the first resonance $k_0=1$ and assume $\kappa\ll 1$. \\
The time derivative of the total energy is
\begin{align}
\frac{dE_{\rm spin}}{dt}=\mathcal{P}_{\rm drive}+\mathcal{P}_{\rm diss}^{(2)}\,,\label{eq.EnergyBalance}
\end{align}
with $\mathcal{P}_{\rm drive}=-\dot{q}n_0\,,\label{eq.Pdrive}$ and $\mathcal{P}_{\rm diss}^{(2)}=-\frac{\hbar}{2}\beta_2 n_0(1-n_0)\dot{\theta}^2$. We introduce $\tilde{n}_0$, the component of $n_0$ oscillating at $\sim \omega$. The product $\dot{q} \tilde{n}_0$ does not vanish after taking the time-average in the expression for $\mathcal{P}_{\rm drive}$.

From Eq.\,(\ref{eq:JA}), the $k=0$ component of $\sin\theta$ oscillating at $\sim \omega$ is
$\widetilde{\sin\theta}=-\cos(\omega t+\varphi_{\rm mod}-\phi)$. The amplitude of other sideband near-resonant with the drive [term $k=2$ in Eq.\,(\ref{eq:JA})] is negligible in the limit $\kappa\ll1$. Using $\tilde{n}_0=\mathcal{O}(U_s/q_0)\ll1$ to simplify Eq.\,(\ref{eq.SpinMixEq1}), we find
\begin{align}
\tilde{n}_0=-\frac{2U_s}{\hbar \omega}\overline{n}_0(1-\overline{n}_0)\sin(\omega t +\varphi_{\rm mod}-\phi)\,.
\end{align} 
Using $\kappa\simeq\Delta q/(\hbar\omega)$ (true if $\kappa\ll1$), the average power delivered by the drive is finally
\begin{align}
\overline{\mathcal{P}}_{\rm drive}=-\omega\kappa U_s\overline{n}_0(1-\overline{n}_0)\sin\phi\,.
\end{align}
When there is no dissipation, this expression can be written as $\overline{\mathcal{P}}_{\rm drive}=-\hbar\omega\dot{\overline{n}}_0/2$. This result has a microscopic interpretation if one treats the driving field as a quantized electromagnetic field. One photon is absorbed to promote a pair of atoms in the $m_F=0$ state to a pair with one atom in $m_F=+1$ and another in $m_F=-1$. The energy in the field is, up to a constant, $E_{\rm field}=N\hbar\omega n_0/2\,$, and $\overline{\mathcal{P}}_{\rm drive}$ correspond to the energy transferred back and forth from the field to the atoms. Alternatively, one could rewrite eq.\,(\ref{eq.EnergyBalance}) as $N\overline{E}_{\rm spin}+\overline{E}_{\rm field}={\rm cste}$.

With dissipation, the system relaxes to the fixed point $S_+$ or to $S_0$. The second case is trivial, since the drive and dissipated power both vanish and nothing happens. Let us discuss the first case. At the fixed points $S_{+}$, the atomic phase is locked to the drive, $i.e.$ $\dot{\theta} \approx -\omega$ and $\overline{\mathcal{P}}_{\rm diss}^{(2)}\approx-\frac{\hbar\omega^2}{2}\beta_2\overline{n}_0(1-\overline{n}_0)\,.$ The energy balance can be rewritten as
\begin{align}
\left.\frac{dE_{\rm spin}}{dt}\right\vert_{S_+}\approx-\omega \overline{n}_0(1-\overline{n}_0) \left[\kappa U_s \sin\phi_+ + \frac{\beta_2 \hbar\omega}{2}\right]\,,\label{eq.EnergyBalanceS+}
\end{align}
The terms in brackets in the right hand side of Eq.\,(\ref{eq.EnergyBalanceS+}) vanishes exactly, as the secular phase takes the value $\sin\phi_+=-\beta_2\hbar\omega/(2\kappa U_s)$ at $S_+$. At the fixed point, the phase lag between the atomic phase and the drive is therefore such that the power delivered by the drive exactly compensates for the energy dissipation.

\section{Quantum treatment of the modulated SMA Hamiltonian}
\label{app:quantum}
We start from the SMA Hamiltonian in Eq.\,(\ref{eq:HSMA}), which we rewrite as
\begin{align} \nonumber
\hat{H}_{\textrm{spin}} &=-q(t) \hat{N}_0 + \frac{U_s}{2N} \left( \hat{V}+\hat{W}+\hat{W}^\dagger\right).
\end{align}
We defined the operators $\hat{V} = \hat{S}_z^2 + 2 \hat{N}_0 ( N - \hat{N}_0)$ and $\hat{W} = 2(\hat{a}_0^\dagger )^2 \hat{a}_{+1} \hat{a}_{-1}$. 
Applying the unitary transformation
\begin{align}
\hat{U}_1 &  = e^{-i\int_0^t \frac{q(t')dt'}{\hbar} \hat{N}_0} =e^{-i p \hat{N}_0},
\end{align}
the transformed Hamiltonian $\hat{H}' = \hat{U}_1\hat{H}\hat{U}_1^\dagger+i\hbar\frac{d \hat{U}_1}{dt}\hat{U}_1^\dagger$ reads
\begin{align}
\hat{H}_1&=\frac{U_s}{2N} \left[ \hat{V}+ \hat{U}_1\left(\hat{W}+\hat{W}^\dagger\right)\hat{U}_1^\dagger\right].
\end{align}
We introduce the Fock basis $\vert N_0, M_z \rangle$ with $N_{\pm 1}=(N-N_0\pm M_z)/2$. The operators $\hat{W}$ (respectively $\hat{W}^\dagger$) only couples states with $M_z=M_z'$ and $N_0=N_0'+2$ (resp. $N_0=N_0'-2$). As a result, the matrix elements of $\hat{U}_1\hat{W}\hat{U}_1^\dagger$ in the Fock basis are the same as the ones of $e^{-2ip}\hat{W}$, implying the equality of both operators.

We now derive an effective Hamiltonian describing the slow secular dynamics. We proceed as in Section\,\ref{app:secular}, using the Jacobi-Anger expansion to rewrite the phase factors and taking the time average over one period of the modulation assuming small detuning $\delta$. We obtain an effective time-averaged Hamiltonian,
\begin{align}
\overline{\hat{H}_1}&=\frac{U_s}{2N}  \hat{V}+ \frac{ \kappa U_s}{2N} \left( e^{i \zeta(t)}\hat{W}+e^{-i \zeta(t)}\hat{W}^\dagger \right).
\end{align}
We finish the calculation with a second unitary transformation $\hat{U}_2  = e^{-i \frac{\zeta(t)}{2} \hat{N}_0}$ to obtain an effective time-independent Hamiltonian
\begin{align}
\hat{H}_{\textrm{eff}}&=-\frac{\hbar \delta}{2} \hat{N}_0+\frac{U_s}{2N}  \hat{V}+ \frac{ \kappa U_s}{2N} \left( \hat{W}+\hat{W}^\dagger \right).
\end{align}
With a mean-field \textit{ansatz} for the many-body spin state, we obtain from this effective Hamiltonian the same secular energy $E_{\rm sec}$ [Eq.\,(\ref{eq:Esec})] as in the classical treatment, \textit{i.e.} mean-field approximation and time averaging can be done in any order.

\section{Stability of the stationary solutions of dissipative model 2.}
\label{app:stabilityfixed}

\subsection{Stability of the fixed points $S_{\pm}$}

To discuss the stability of the fixed points $S_\pm$, we linearise Eqs.\,(\ref{eq.TimeAvgDissEq1},\ref{eq.TimeAvgEq2}) using $\overline{n_0}=\overline{n}_{0,\pm}+\delta\overline{n}_{0,\pm}$ and $\phi=\phi_\pm + \delta\phi_\pm$. We find
\begin{align}
\hbar 
\begin{pmatrix}
\delta\dot{\overline{n}}_{0,\pm} \\
{\delta\dot{\phi}}_\pm
\end{pmatrix}
&= M_\pm 
\begin{pmatrix}
\delta\overline{n}_{0,\pm} \\
{\delta\phi}_\pm
\end{pmatrix}\,\\
M_\pm &= 
\begin{pmatrix}
0 & \pm 2\kappa U_s n_{0,\pm}(1-n_{0,\pm})\cos\epsilon\\
-2\hbar\delta_\pm & -2\kappa U_s\frac{\delta}{\delta_\pm}\sin\epsilon	
\end{pmatrix}\,\nonumber
\end{align}
The solutions are stable if the eigenvalues of the matrices $M_\pm$ have negative real parts. For simplicity, we consider the situation $\vert\sin\epsilon\vert=\beta_2\hbar\omega/(2\kappa U_s)\ll1$. One can show that the results below hold as long as $\beta_2\hbar\omega/(2\kappa U_s)<1$, the same condition as for the existence of the fixed points.

In the limit $\epsilon \ll 1$, the eigenvalues of $M_+$ are approximately given by $X_{+,1} \simeq \beta_2\hbar\omega\frac{\delta}{2\delta_+}+i\sqrt{\Delta}\,,$ and $X_{+,2} = X_{+,1}^*\,,$ with $\Delta=8\overline{n}_{0,+}(1-\overline{n}_{0,+})\kappa(1+\kappa)U_s^2$. Therefore, $S_+$ is stable for $\delta<0$, and unstable otherwise. Turning to $S_-$, the eigenvalues are $X_{-,1}\simeq\sqrt{\Delta}$ and $X_{-,2} \simeq -X_{-,1}$ to leading order in $\beta_2$, and $S_-$ is therefore always unstable. Note that our conclusions are established for the experimentally relevant case $0 \leq \kappa < 1$. The roles of $S_\pm$ would be reversed for $\kappa <0$. 

\subsection{Stability of the limit cycles $S_{0,1}$}
\label{app:stabilitylimit}

We focus first on $S_1$. We consider small deviations, \textit{i.e.} $\overline{n}_0=1-\epsilon$ and linearize Eqs.\,(\ref{eq.TimeAvgDissEq1},\ref{eq.TimeAvgEq2}) to the lowest order in $\epsilon$,
\begin{align}
-\hbar\dot{\epsilon}&=2\kappa U_s\sin\phi\epsilon+2\beta_2q_0\epsilon\,,\label{eq.n0linS1}\\
\hbar\dot{\phi}&=-\hbar\delta-2U_s(1+\kappa\cos\phi)\,\label{eq.philinS1}.
\end{align} 
We integrate Eq.\,(\ref{eq.n0linS1}),
\begin{align}
\big[\ln\epsilon\big]_{\epsilon(0)}^{\epsilon(t)}=-\frac{2\kappa U_s}{\hbar}\int_0^t \sin\phi(t') dt' - \frac{2\beta_2q_0t}{\hbar}.\nonumber
\end{align} 
Making the change of variable $t\to\phi$ and using Eq.\,(\ref{eq.philinS1}), we find
\begin{align}
\epsilon(t)=\epsilon(0)e^{-4t/\tau_2}\frac{1+a_1\cos\phi(0)}{1+a_1\cos\phi(t)},
\end{align}
with $a_1=2\kappa U_s/[2U_s+\hbar\delta]$ and $\tau_2=2\hbar/(\beta_2 q_0)$. As $\phi$ is running, $\epsilon(t)$ diverges iif $\vert a_1\vert >1$. This defines the instability region of $S_1$ as $\delta\in[-2U_s(1+\kappa),-2U_s(1-\kappa)]$. Remarkably, this results is independent of the precise value of $\beta_2$ as long as it is strictly positive. 

A similar calculation for $S_0$ with $\epsilon=\overline{n}_0$ yields 
\begin{align}
\epsilon(t)=\epsilon(0)e^{4t/\tau_2}\frac{1+a_0\cos\phi(0)}{1+a_0\cos\phi(t)},
\end{align}
with $a_0=2\kappa U_s/[2U_s-\hbar\delta]$. Since $\beta_2>0$, we find that $S_0$ is always unstable.

\bibliography{shapiroBib}
\bibliographystyle{apsrev_nourl}

\end{document}